\documentclass[twocolumn,aps,pre,showpacs,floatfix]{revtex4}
\usepackage[pdfborder=0 0 0]{hyperref}
\usepackage[dvips]{graphicx}
\usepackage{amsmath}
\usepackage{color}
\usepackage{amsfonts}
\usepackage{amssymb}
\usepackage{epstopdf}
\DeclareGraphicsExtensions{.eps,.png,.pdf}
\begin{document}

\markboth{Sethia}{Sethia}
\title{Existence and stability of traveling wave states in a ring of non-locally coupled phase oscillators
with propagation delays}
\author{Gautam C Sethia}
\email{gautam@ipr.res.in}
\author{Abhijit Sen}
\email{abhijit@ipr.res.in}
\affiliation{Institute for Plasma Research, Bhat, Gandhinagar 382 428, India}
\date{\today}

\pacs{05.45.Ra, 05.45.Xt, 89.75.-k}%

\begin{abstract}
We investigate the existence and stability of traveling wave solutions in a continuum field of non-locally coupled
identical phase oscillators with distance-dependent propagation delays. A comprehensive stability diagram in the 
parametric space of the system is presented that shows a rich structure of multi-stable regions and illuminates
the relative influences of time delay, the non-locality parameter and the intrinsic oscillator frequency on the dynamics of 
these states. A decrease in the intrinsic oscillator frequency leads to a break-up of the stability domains of the traveling waves  
into  disconnected  regions in the parametric space. These regions exhibit a tongue structure for high connectivity whereas 
they submerge into the stable region of the synchronous state for low connectivity. A novel finding is the existence of forbidden regions
in the parametric space where no phase-locked solutions are possible.  We also discover a new class of non-stationary \textit{breather} states 
for this model system that are characterized by periodic oscillations of the complex order parameter. 
\end{abstract}

\keywords{synchronization, propagation delays, non-local coupling, phase oscillators, marginal stability curve}

\pacs{05.45.Ra, 05.45.Xt, 89.75.-k}
 
\maketitle
\section{Introduction}

The classic Kuramoto model \cite{kuramoto75}, consisting of a ring of phase coupled oscillators, has for long served as a very useful and convenient paradigm for investigating collective phenomena in a variety of physical, chemical and biological systems \cite{Kuramoto84book,buck88,wiesenfeld96,pikovsky}.   The original model employing a simple global (all-to-all) and constant uniform coupling between the oscillators, has subsequently been extended and generalized in various ways to improve its applicability as well as to explore a wider range of dynamics \cite{ermentrout90a,crook97,yeung99,zanette00,kuramoto02,kuramoto03,earl03,ko07,sethia08,sethia10}.  One such generalized version that provides a closer representation of many real life situations, employs non-local coupling to account for diminution in the coupling strength as a function of distance and distance dependent time-delayed coupling to account for finite propagation velocities of information signals exchanged between oscillators \cite{sethia10}. Such a model displays a much richer variety of collective excitations including synchronous states \cite{sethia10} and exotic {\it chimera} states \cite{sethia08}. Both non-locality and time delay play important roles in influencing the equilibrium and stability properties of these collective states and their combined presence often introduces novel effects. To the best of our knowledge,  an analysis of the equilibria and stability of traveling wave states for the generalized model has not been done so far, excepting a few past studies that have employed simplified models or looked at special limits. Crook et al.~\cite{crook97} considered a continuum system of coupled identical oscillators with a spatially decaying interaction kernel and modeled the space dependent time delay contribution through an effective phase shift term in the interaction. They also took the size of the system to be infinite for mathematical convenience but thereby effectively reduced the nature of the mutual coupling to a ``local'' one (since the interaction length is always much smaller than the system length). Zanette \cite{zanette00} studied another simplified version of the generalized system where he adopted a {distance} {independent} (global) coupling between the oscillators positioned along a ring but introduced a {distance} {dependent} time delay in the interaction. He obtained numerical results on the stability of synchronous and propagating traveling waves and also some analytic results in the limit of small delay. Ko \& Ermentrout \cite{ko07} recently investigated the effects of distance dependent delays in sparsely connected oscillator systems and found that a small fraction of connections with time delay can destabilize the synchronous states. Since traveling wave states are basic ingredients for pattern formation in many real life systems such as arrays of  
Josephson junctions \cite{phillips93}, chemical oscillators \cite{kuramoto02,shima04}, neural networks producing snail shell
patterns and ocular dominance stripes \cite{ermentrout86} etc. a determination of their stability properties within the framework of the generalized model is of fundamental importance. 

In this paper we present a systematic and comprehensive stability diagram of traveling wave states adopting a suite of numerical approaches that has proved successful in the past in delineating the stability of synchronous states of this fully generalized model \cite{sethia10}.  The stability diagram shows a rich structure of multi-stable regions in the parametric space of time delay and non-locality of the coupling with the traveling waves losing their stability across the marginal curves through a Hopf bifurcation. Interspersed among the stable regions there exist \textit{forbidden} regions (whose size and shape show a strong dependence on the intrinsic frequency of the oscillator) in the phase diagram where no stable phase-locked solutions are possible. Our numerical investigations also reveal the existence of a new class of non-stationary states for this system that are characterized by periodic oscillations of the complex order parameter and for which the phase and its temporal derivative have a non-linear dependence on space.  Finally in support of our detailed numerical results we also propose an analytic heuristic necessary condition for stability that provides an upper as well as a lower stability limit for the traveling wave  solutions.  
  
\section{Model system and its traveling wave states}

Our dimensionless model equation representing the dynamics of a continuum of coupled identical phase oscillators
located on a ring, $x\in\lbrack-L,L]$ is given by \cite{sethia10},

\begin{equation}
\frac{\partial \phi(x,t) }{\partial t} =\omega-\int_{-1}^{1
}G(z)\sin\left[  \phi(x,t)-\phi\left( x-z, t-|z|\tau_{m}\right)  \right]dz. 
\label{phase}
\end{equation}

\noindent
where $\phi(x,t) \in [0,2\pi)$ is the phase of the oscillator located at $x$ and at time $t$ and  whose intrinsic oscillation frequency is $\omega>0$.  In (\ref{phase}) the space co-ordinates $x$ and $z$ have been normalized by $L$ and time and frequencies have been made dimensionless
by the prescription $t = K t$, $\omega = \omega /K$ where $K$ is the strength of the coupling between the oscillators. The quantity $\tau_m = 1/v$ denotes the maximum time delay in the system with $v$ representing the signal propagation speed. The normalized function $G:[-1,1]\rightarrow\mathbb{R}$  describes the coupling kernel and is an even function of the form,
 \begin{equation}
G(z)=\dfrac{\kappa}{2(1.0-e^{-\kappa})}e^{-\kappa|z|} 
\label{kernel} 
\end{equation} 
\noindent
where $\kappa$ is a dimensionless quantity denoting the inverse of the interaction scale length and is a measure of
the non-locality of the coupling. 

Before any further deliberations it is convenient to define a mean delay parameter by
\begin{equation}
\bar{\tau}=\int_{-1}^{1}G(z)\tau_{m}|z|\, dz\
\label{mtaueqn}
\end{equation}
which weights the individual delays with the corresponding connection
weights and for the exponential connectivity given by Eq.~(\ref{kernel}), becomes 
\begin{equation}
\bar{\tau}=\tau_{m}/c_{\kappa}
\end{equation}
\label{mtau}
where $c_\kappa = \frac{\kappa(e^\kappa - 1)}{e^\kappa - 1 - \kappa}.$

We look for phase-locked solutions of Eq.~(\ref{phase}) of the form 
\begin{equation}
\phi_{\Omega,k}(x,t)=\Omega t+\pi k x+\phi_0.
\label{sol}
\end{equation} 
These solutions are \textit{phase-locked} in the sense that the difference in phases at two fixed locations in space does not change with time. They describe both the synchronous ($\Omega \neq 0,k=0$) as well as the traveling wave solutions ($\Omega\neq0,k\neq0$).  The traveling wave solution are thus phase-locked solutions  in which each oscillator has the same frequency $\Omega$ but the phase varies monotonically along the ring. Note that $k$ is an integer due to periodic boundary conditions, and the value of $\phi_0$ can be taken to be zero by a translation. 
Substituting Eq.~(\ref{sol}) into Eq.~(\ref{phase}) gives the following dispersion relation,
\begin{widetext}
\begin{align}
\Omega =\omega-&\int_{-1}^{1}G(z)\sin\left[  \Omega \tau_{m} |z|+\pi kz 
\right]  \,dz, \nonumber \\
=\omega-&\frac{e^{\kappa } \kappa  \Omega \tau_{m}  \left(-k^2 \pi ^2+\kappa ^2+\Omega^2  \tau_{m}^2\right)}{\left(-1+e^{\kappa }\right) \left(\kappa ^2+(k \pi -\Omega \tau_{m} )^2\right) \left(\kappa ^2+(k \pi +\Omega \tau_{m} )^2\right)}\nonumber\\
-&\frac{(-1)^k \kappa  \left(-\Omega \tau_{m}  \left(-k^2 \pi ^2+\kappa ^2+\Omega^2  \tau_{m}^2\right) \text{Cos}[\Omega \tau_{m} ]-\kappa  \left(k^2 \pi ^2+\kappa ^2+\Omega^2  \tau_{m}^2\right) \text{Sin}[\Omega \tau_{m} ]\right)}{\left(-1+e^{\kappa }\right) \left(\kappa ^2+(k \pi -\Omega \tau_{m} )^2\right) \left(\kappa ^2+(k \pi +\Omega \tau_{m} )^2\right)}
\label{eqlbm}
\end{align}
\end{widetext}
The solutions of the above general dispersion relation represent the frequencies of various collective states of the system
and its transcendental nature implies that $\Omega$ can be multi-valued in principle for a given set of parameters 
$\omega, \tau_{m}$, $k$ and $\kappa$. It is interesting to take various limits of Eq.~(\ref{eqlbm}) and relate them to some of the past
work on simplified models. In the limit when $\kappa \rightarrow 0$,  Eq.~(\ref{eqlbm}) reduces to,
\begin{equation}
\Omega = \omega - \frac{(1 - (-1)^{k}cos(\Omega \tau_m))\Omega \tau_m}{-(k\pi)^{2}+(\Omega \tau_m)^{2} }
\label{zanette}
\end{equation}
This is the reduced model that Zanette \cite{zanette00} had investigated to obtain propagating waves in a system of globally  coupled oscillators through the
introduction of distance dependent delays.  The other interesting scenario is that of  $\kappa \gg 1$, which corresponds to the situation of local coupling among the oscillators. In this case Eq. (\ref{eqlbm}) reduces to,
\begin{equation}
\Omega=\omega-\dfrac{\kappa}{2}\left(\dfrac{-k \pi+\Omega \tau_m}{\kappa^{2}+(k \pi-\Omega \tau_m)^{2}}+\dfrac{k \pi+\Omega \tau_m}{\kappa^{2}+(k \pi+\Omega \tau_m)^{2}}\right)
\label{crook}
\end{equation}
The local limit can also be approached by taking $L \rightarrow \infty$ in the unscaled form of the model and hence of Eq.~(\ref{eqlbm}) in which $\tau_m$ is replaced by $L/v$ and $\kappa$ by $\kappa L$. The mean delay ($\bar{\tau}_{\infty}$) in this system equals $\frac{1}{\kappa v}$ and Eq.~(\ref{eqlbm}) can be rewritten in terms of   $\bar{\tau}_{\infty}$ as
\begin{equation}
\Omega=\omega-\dfrac{1}{2}\left(\dfrac{-p+\Omega \bar{\tau}_{\infty}}{1+(p-\Omega \bar{\tau}_{\infty})^{2}}+\dfrac{p+\Omega \bar{\tau}_{\infty}}{1+(p+\Omega \bar{\tau}_{\infty})^{2}}\right)
\label{crook1}
\end{equation}
where $p$ is a real number denoting the wave number for the infinite system. This is a third-order polynomial equation in $\Omega$ and can at most have three real solutions in contrast to the higher number of multiple roots of the transcendental Eq.~(\ref{eqlbm}). Crook et al. \cite{crook97} further simplified the model by replacing the explicit delay time dependence by  a space-dependent phase shift in the coupling term and thereby Eq.~(\ref{crook1}) further reduces  to   
\begin{equation}
\Omega=\omega-\dfrac{1}{2}\left(\dfrac{-p+ \bar{\tau}_{\infty}}{1+(p- \bar{\tau}_{\infty})^{2}}+\dfrac{p+\bar{\tau}_{\infty}}{1+(p+ \bar{\tau}_{\infty})^{2}}\right)
\label{crook2}
\end{equation}
In this approximation, $\Omega$ is simply rescaled by the value of $\omega$ and was taken to be zero by Crook et al.\cite{crook97} as $\omega$ played no role in the stability of the system under this approximation.
We now turn to the generalized dispersion relation Eq.~(\ref{eqlbm}) and discuss
its solutions. 

\begin{figure}
\includegraphics[width=0.45\textwidth]{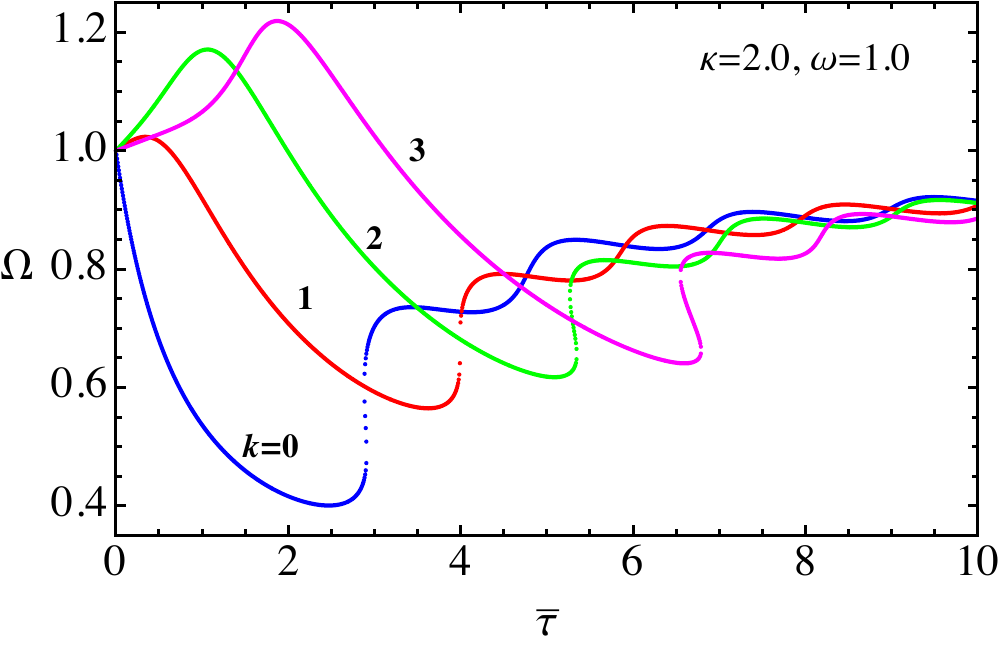}
\caption{(Color online) Synchronous frequency $\Omega$ for various wave mode numbers $k$ for fixed values of $\kappa$ and $\omega$. $k=0,1,2$ and $3$ are shown in blue,  red,  green and in magenta respectively and are also marked with the corresponding $k$ values.}
\label{fig:fig1}
\end{figure}

\begin{figure}
\begin{tabular}
[c]{c}
\includegraphics[width=0.45\textwidth]{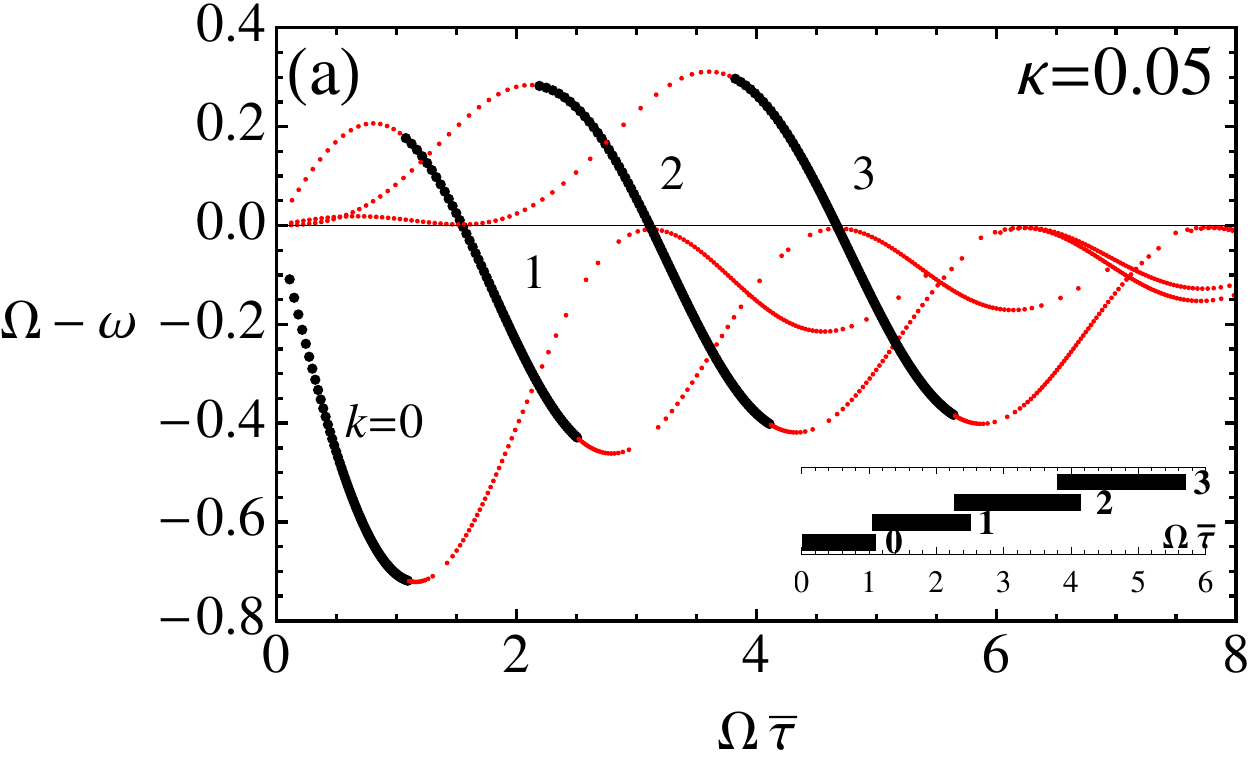}\\
\includegraphics[width=0.45\textwidth]{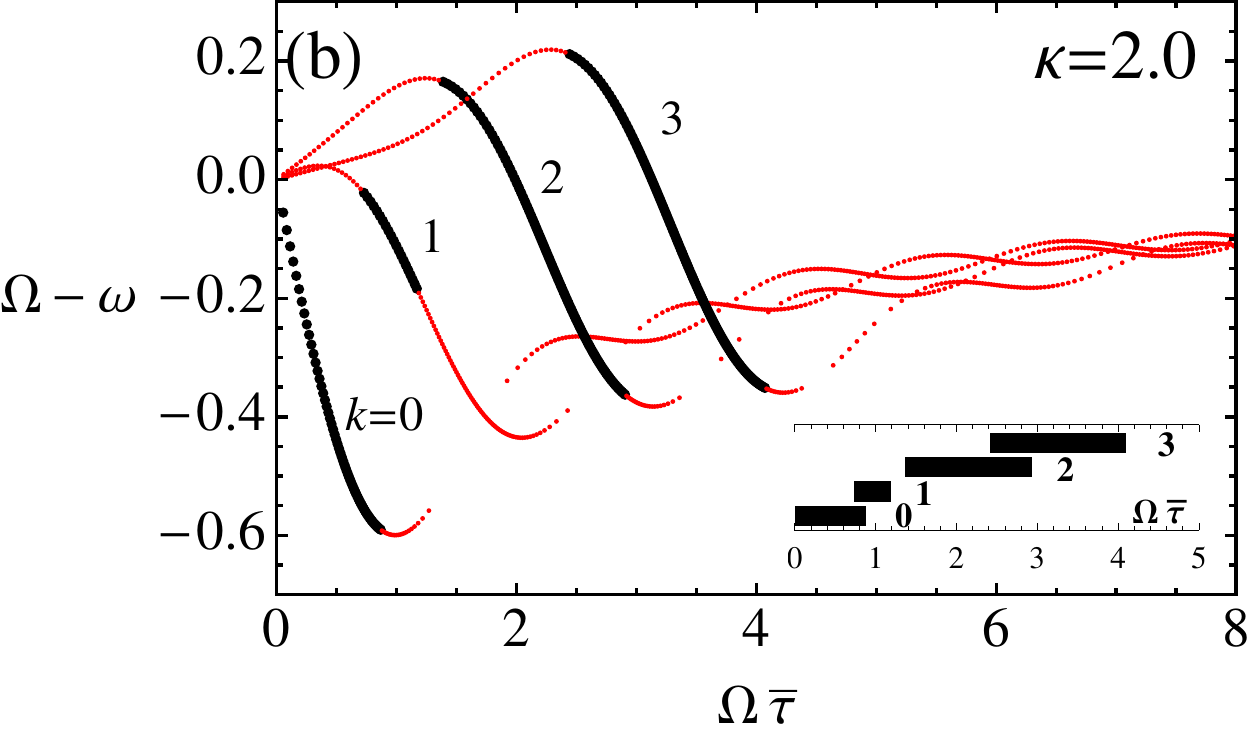}\\
\includegraphics[width=0.45\textwidth]{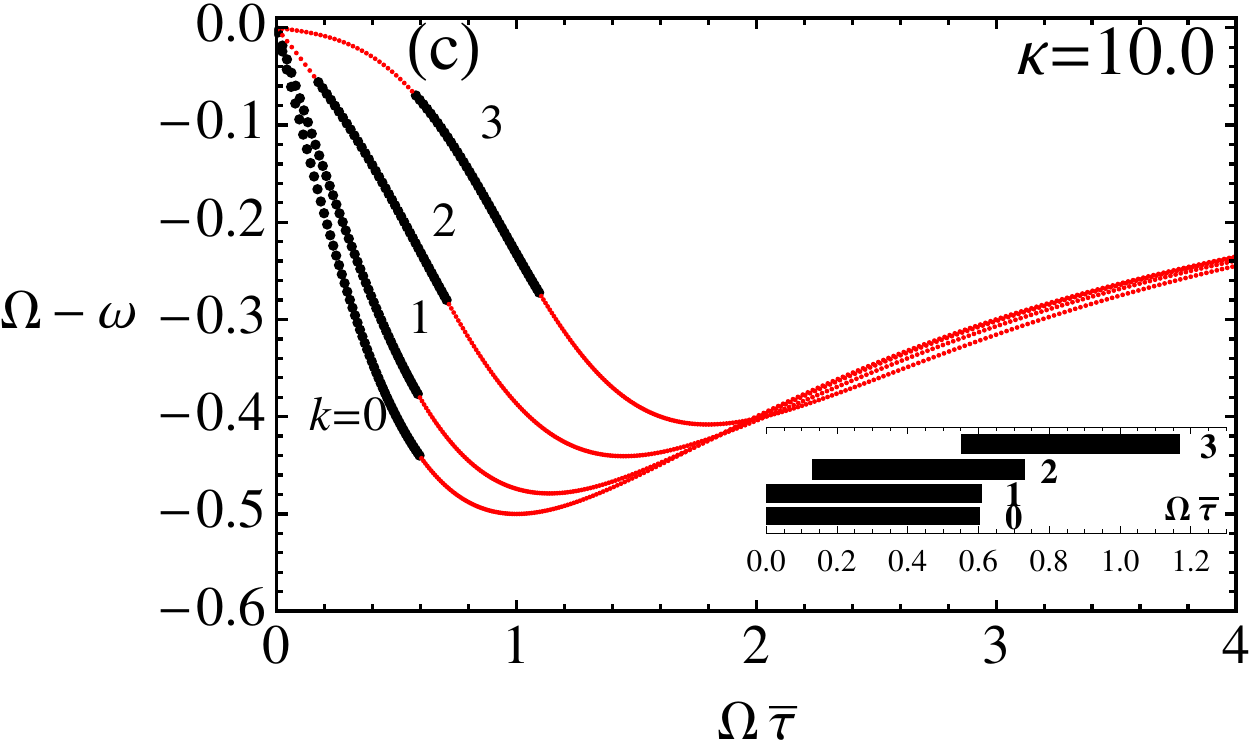}
\end{tabular}
\caption{(Color online) Plots of $\Omega-\omega$ vs $\Omega \bar{\tau}$ for various phase-locked states ($k=0,1,2,3$) at three different values of $\kappa$. The shapes of all the curves are invariant to $\omega$ but the stable region  is $\omega$ dependent. The solid black segments denote stable states for $\omega=1$. The thin segments shown in red denote the unstable regions. The insets highlight the width of stable regimes for different $k$ values.}
\label{fig:fig2}
\end{figure}

The dispersion relation given by Eq.~(\ref{eqlbm}) can be recast in the form :
\begin{equation}
 \Omega-\omega = H(\Omega \bar{\tau},\kappa)
\label{H}
\end{equation} 
where
\begin{equation}
H(\Omega \bar{\tau},\kappa) = -\int_{-1}^{1}G(z)\sin\left( c_\kappa \Omega \bar\tau  |z|+\pi k z\right) dz 
\label{H_def}
\end{equation} 
Fig.~\ref{fig:fig1} plots the numerical solutions $\Omega$ of Eq.~(\ref{H}) as a function of $\bar{\tau}$ for $\kappa=2.0$ \& $\omega=1.0$ for different values of the wave number $k$. 
We note from Eq.~(\ref{H}) that the difference between the phase-locked frequency ($\Omega$) and the intrinsic oscillator frequency ($\omega$) is a function of the corresponding $\Omega\bar{\tau}$ and the non-locality parameter $\kappa$  and has no explicit dependence on $\omega$.  This provides some simplification in the sense that if we plot the solutions of Eq.~(\ref{H}) in the phase space of $\Omega-\omega$ versus $\Omega\bar{\tau}$  or equivalently in the phase space of $H(\Omega\bar{\tau},\kappa)$ versus $\Omega\bar{\tau}$, the results hold for any $\omega$.  Figure ~\ref{fig:fig2} plots these universal curves which are the solutions of Eq.~(\ref{H}) for traveling waves with $k=1,2$ and $3$ respectively and for three different values of $\kappa$. For completeness we have also included the results for the synchronous state ($k=0$) that were previously obtained in \cite{sethia10}. The three different values of $\kappa$ have been chosen to probe parametric domains  in the vicinities of global coupling ($\kappa=0.05$,  Fig.~\ref{fig:fig2}(a)), intermediate (or non-local) coupling ($\kappa=2.0$, Fig.~\ref{fig:fig2}(b)) and  local coupling ($\kappa=10.0$, Fig.~\ref{fig:fig2}(c)). While the equilibrium curves are universal (i.e. independent of $\omega$) the stability domains of these solutions indicated by the solid portions of the curves do depend on $\omega$. In Fig.~\ref{fig:fig2} the solid portions of the curves in all the three panels denote stable states for $\omega=1$. The insets in each panel indicate the widths of the stable regions in $\Omega\bar{\tau}$ space for the corresponding  $k$ values. In the next section, we discuss  the method by which we determine these stable  regions. 
\section{stability of the phase-locked states}
The linear stability of the solutions $\phi_{\Omega,k}(x,t)$ of Eq.~(\ref{phase}) 
is determined by the variational equation
\begin{align}
\frac{\partial}{\partial t}u(x,t)  = &-\int_{-1}^{1}G(z)\cos\left[c_{\kappa}\Omega\bar{\tau} |z|+\pi k z \right]\nonumber\\
& \times [u(x,t)-u(x-z,t-|z|c_{\kappa}\bar{\tau})]dz
\label{veqn}
\end{align}
where $u(x,t)=\phi(x,t)-\phi_{\Omega,k}(x,t)$. With the ansatz
$u(x,t) \sim e^{\lambda t}e^{i\pi nx}$, $\lambda\in\mathbb{C}$, $n\in\mathbb{Z}$, we
obtain the eigenvalue equation:
\begin{align}
f(\lambda)\equiv\lambda & +\int_{-1}^{1}G(z)\cos\left(  c_{\kappa}\Omega\bar{\tau}|z|+\pi k z\right) \nonumber\\
& \times \left(  1-e^{-\lambda|z|c_{\kappa}\bar{\tau}}e^{-i\pi n z}\right)  \,dz=0
\label{delay-lambda}
\end{align}

Writing $\lambda = \lambda_R + i\lambda_I$ and separating Eq.~(\ref{delay-lambda}) into its real and imaginary parts we get
\begin{align}
\lambda_R  =& -\int_{-1}^{1}G(z)\cos\left(   c_{\kappa}\Omega \bar{\tau}|z|+\pi k z\right)\nonumber\\
& \times\left[  1-e^{-\lambda_R|z|c_{\kappa}\bar{\tau}}\cos\left( \lambda_I |z|c_{\kappa}\bar{\tau} + \pi nz\right)\right]dz,\label{lambdaR}\\
\lambda_I  =& -\int_{-1}^{1}G(z)\cos\left(   c_{\kappa}\Omega\bar{\tau}|z|+\pi k z\right)\nonumber\\
& \times e^{-\lambda_R|z|c_{\kappa}\bar{\tau}}\sin\left(\lambda_I|z|c_{\kappa}\bar{\tau} + \pi nz\right)dz.\label{lambdaI}
\end{align}
The linear stability of the traveling wave state requires that all solutions of Eq.~(\ref{delay-lambda}) have $\lambda_R < 0$ for all non-zero integer values of $n$. 
The implicit and transcendental nature of the eigenvalue equation makes it difficult to obtain an analytic solution so we adopt a multi-pronged numerical approach to determine
the marginal stability curves. 
For a given phase-locked state (i.e. fixing a value of $k$ to 1, 2,3 etc) and a fixed intrinsic oscillator frequency $\omega$, we solve Eq.~(\ref{H}) for $\Omega$ for a range of $\kappa$ values to span through the whole range of global to local coupling. We vary $\kappa$ from $10^{-6}$ to $15$. The solutions so obtained are tested  for eigenvalues with positive real parts by the Cauchy's argument principle which states that, the number of unstable roots $m$ of $f(\lambda)$ is given by :
\begin{equation}
m = \frac{1}{2\pi i}\oint_C\frac{f'(\lambda)}{f(\lambda)}d\lambda
\end{equation}
where the closed contour $C$ encloses a domain in the right half of the complex $\lambda$ plane with the imaginary axis forming its left boundary.
For each of the $\kappa$ values, $\tau_{m}$ (and hence $\bar{\tau}$) is varied from zero to the values where $\lambda_{R}=0$ transitions are  noted. Although this analysis is carried out for a large number of non-zero integer values of $n$, it is found that for synchronous solutions (i.e. $k=0$), the lowest mode number, namely, $n=1$ is the first one to get destabilized \cite{sethia10}  whereas for the higher wave numbers ($k=1,2,3$), the first mode number $n$ to get destabilized is normally less than $10$ for the parameter range covered in the present work. The analysis provides us $\Omega, \tau_{m} $ and $n$ near  $\lambda_{R}=0$ transitions for any given values of $\kappa, \omega$ and $k$. The transition values of $\Omega$ and $\tau_{m}$ are further refined (by using these values as initial guesses) by solving simultaneously a set of three equations (Eqs.\ref{eqlbm},\ref{lambdaR} and \ref{lambdaI}) with $\lambda_{R}=0$ for $\Omega, \tau_{m}$ and $\lambda_{I}$. We further find that $\lambda_{I} \neq 0$ at the marginal point for any of the traveling wave solutions given by $k=1,2,3..$ indicating that they lose their stability through a Hopf bifurcation. This is in contrast to synchronous states ($k=0$)  which were shown to lose their stability through a saddle-node bifurcation with $\lambda_{I} = 0$ \cite{sethia10}.   One common feature that the traveling waves share with the synchronous states is that their stability regions are always
restricted to the lowest branch of the equilibrium solutions where the curves have a negative slope. This is clearly seen in Fig.~\ref{fig:fig2} and suggests that one can write down a heuristic
 necessary condition for the stability of the traveling wave solutions to be $H'<0$, where the prime indicates a derivative of $H(\Omega \bar{\tau},\kappa)$ $w.r.t$ $\Omega \bar{\tau}$.  Using the expression for $H$ given in  (\ref{H_def}) the heuristic condition can be expressed as, 
\begin{equation}
\int_{-1}^{1}|z|G(z)\cos\left( c_\kappa \Omega \bar\tau |z|+\pi k z\right)dz\;>\;0
\label{nec_cond}
\end{equation} 
\begin{figure}
\begin{tabular}[c]{c}
\includegraphics[width=0.45\textwidth]{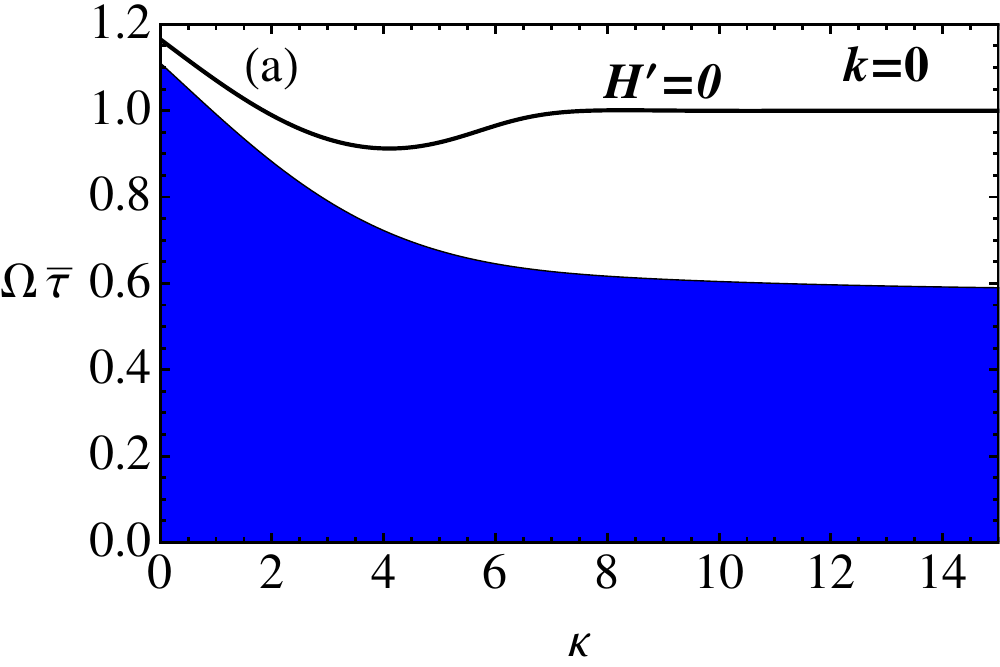}\\
\includegraphics[width=0.45\textwidth]{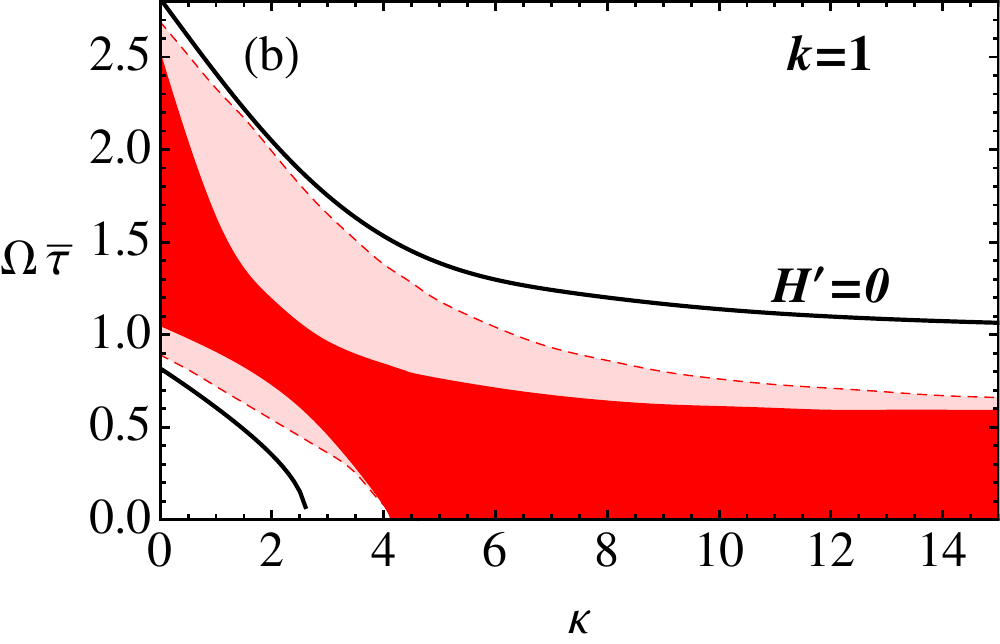}\\
\includegraphics[width=0.45\textwidth]{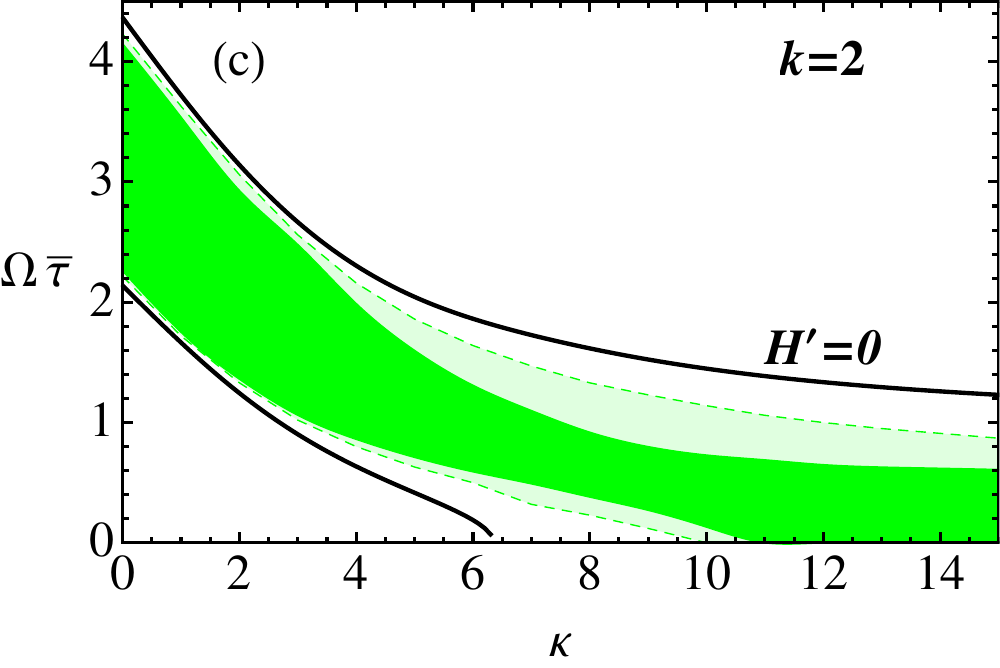}\\
\includegraphics[width=0.45\textwidth]{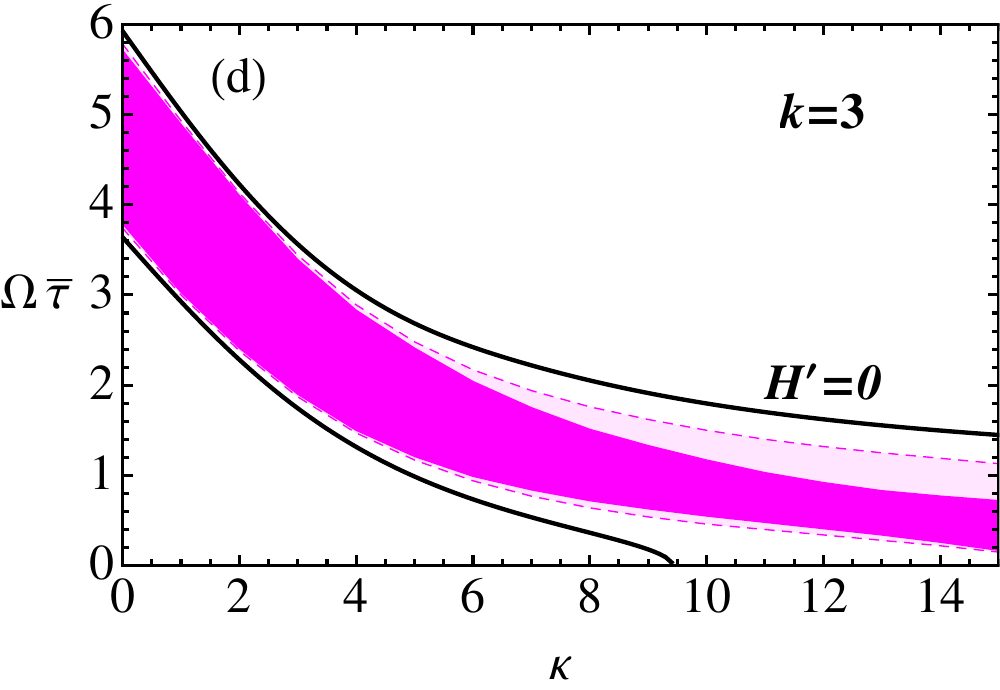}
\end{tabular}
\caption{(Color ) Stability diagrams in the parameter space of $\Omega\bar{\tau} $ vs $\kappa$ for mode numbers $k=0,1,2$ and $3$, shown in blue, red, green and in magenta respectiveley. The solid black curves in each panel represent the condition $H'=0$. Solid color areas  represent stability regions for $\omega=1$ and the lighter shaded areas show the expansion of the stability region when $\omega=10$. Note that the stability region in panel $(a)$ for $k=0$ is independent of $\omega$. 
 }
\label{fig:fig3}
\end{figure}
\begin{figure}
\begin{tabular}[c]{c}
\includegraphics[width=0.45\textwidth]{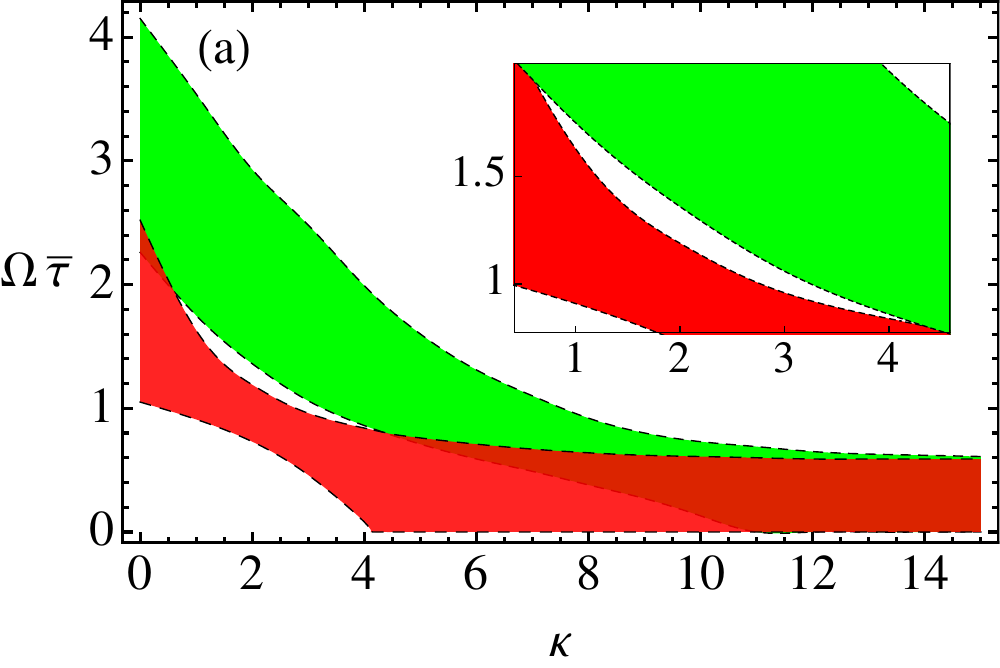}\\
\includegraphics[width=0.45\textwidth]{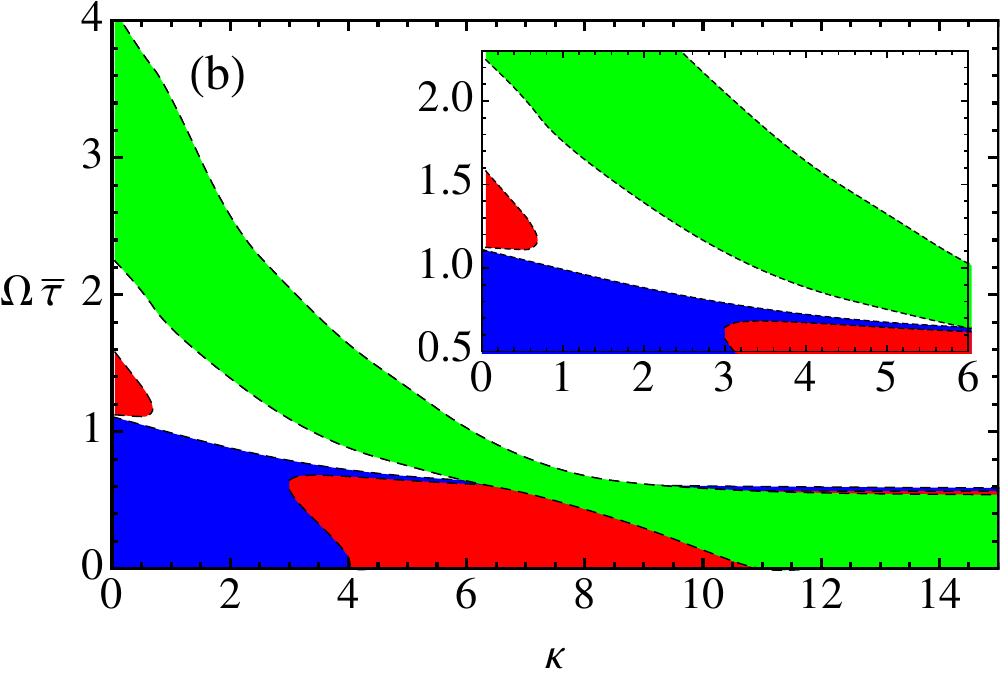}\\
\includegraphics[width=0.45\textwidth]{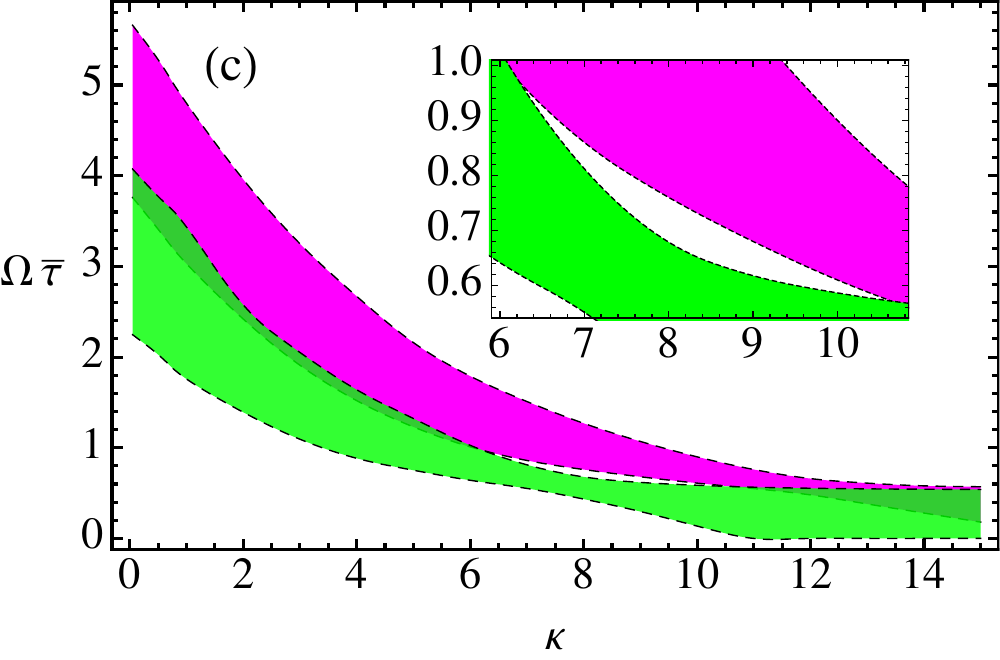}\\
\includegraphics[width=0.45\textwidth]{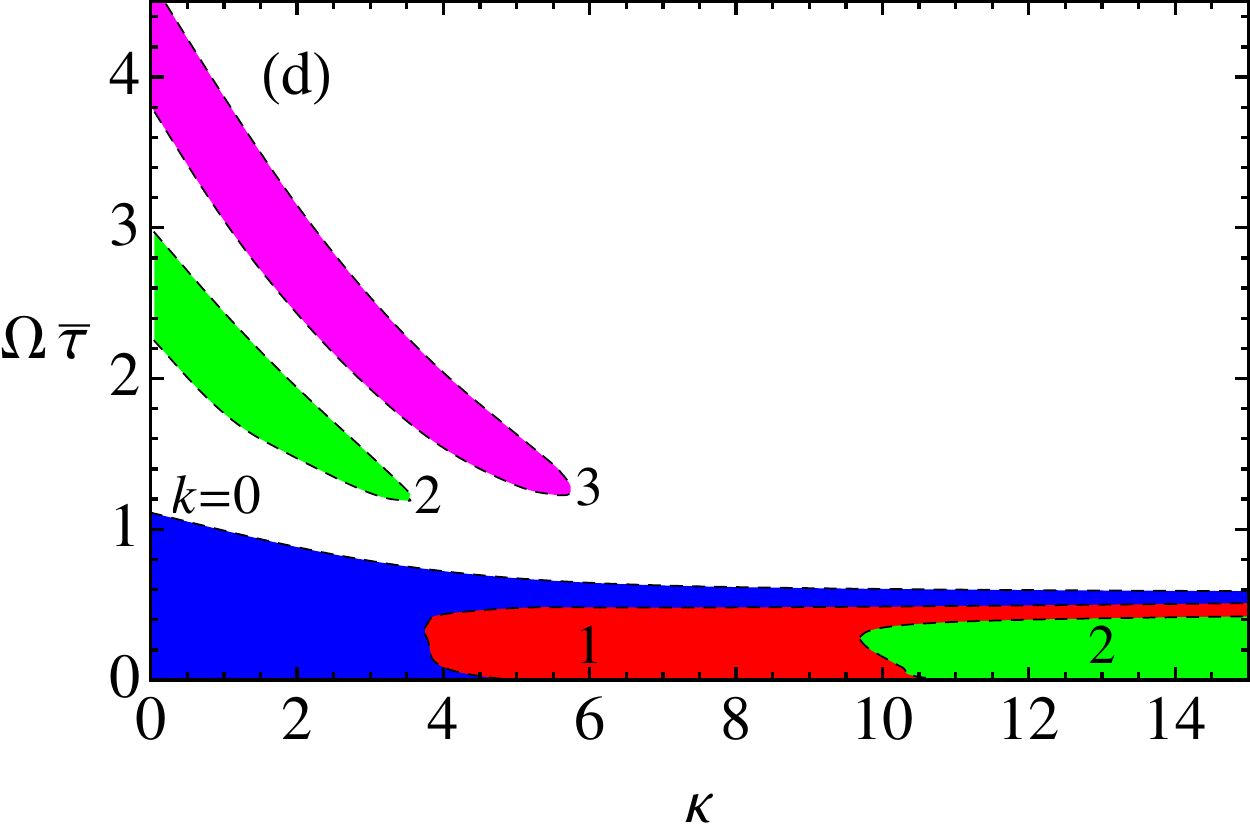}
\end{tabular}
\caption{ (Color) (a) The superimposed stability diagrams of $k=1$ and $2$ for $\omega=1$ shows a forbidden region where no phase-locked states are possible. The inset gives an enlarged view of the forbidden region. Similar forbidden regions are shown for $\omega=0.8$ in $(b)$ \& $(c)$  and for $\omega=0.5$ in $(d)$. The color coding is the same as in Fig.~\ref{fig:fig3}.}
\label{fig:fig4}
\end{figure}
For $k=0$ the above necessary condition applies to the synchronous states as was discussed in \cite{sethia10}. As was also shown in \cite{sethia10}, one can further obtain an
analytic expression for a sufficient condition for stability of the synchronous states given by,
\begin{equation}  
   \int_{-1}^{1}G(z)\cos\left(   c_{\kappa}\Omega\bar{\tau} |z|\right)\left[  1-\cos\left(\pi z\right)\right]dz > 0
\label{marg_stab0}
\end{equation}
This expression is easily obtained from Eq.~(\ref{lambdaR}) by exploiting the facts that the lowest mode number ($n=1$) is the first one to get destabilized and also the fact that the synchronous states lose their stability through a saddle-node bifurcation for which $\lambda_I=0$ \cite{sethia10}. 

The above expression provides simple and handy analytic stability criteria in limiting cases, e.g. $\Omega \bar{\tau}<\frac{\pi}{2\sqrt{2}}$ for global ($\kappa\to 0$) coupling and $\Omega \bar{\tau}<\frac{1}{\sqrt{3}}$  for the local ($\kappa\to \infty$) coupling.  In terms of frequency depression ($\Omega-\omega$), these inequalities for the stability become  $\Omega-\omega > -\frac{\sin ^{2}(\frac{\pi}{2\sqrt{2}})}{\frac{\pi}{2\sqrt{2}}} $ for global and $ \Omega-\omega > -\dfrac{\sqrt{3}}{4} $ for local coupling.  We however note that under the assumption of constant delay $\tau$, the synchronization condition from Eq.~(\ref{marg_stab0}) simply becomes $\Omega\tau<\pi/2$ which agrees with the previously obtained results for constant delay systems \cite{yeung99,earl03}. The stability condition obtained for local coupling ($\Omega \bar{\tau}<\frac{1}{\sqrt{3}}$) matches with that of given by Crook et al.~\cite{crook97}  for infinite system with non-local coupling incorporating propagational delays as phase shifts.

For our present stability studies of traveling waves it is not possible to derive such an analytic expression for a sufficient condition since the transition to an instability is now through a Hopf bifurcation and hence $\lambda_I \neq 0$. Nevertheless the heuristic necessary condition (\ref{nec_cond}) serves a useful purpose by providing an upper and lower limit for the stability domains as we will see soon. 
Fig.~\ref{fig:fig2} also shows some other interesting features, namely, the presence of multi-stable regions where more than one traveling wave modes are stable (seen very prominently in (Fig.~\ref{fig:fig2}(c)) and the existence of a forbidden region (between $k=1$ and $k=2$ for $\kappa=2$ in Fig.~\ref{fig:fig2}(b)) where no phase-locked states can exist. The size and shape of these
regions are further found to depend on the magnitude of the intrinsic frequency $\omega$ and of the non-locality parameter $\kappa$. To investigate this dependence we have examined the stability of the traveling waves for two different values of $\omega$, namely, $\omega=1$ and $\omega =10$ and the results are displayed as a consolidated diagram in the phase space of $\Omega \bar\tau$ versus $\kappa$ in Fig.~\ref{fig:fig3}. In this figure we have also plotted the $H^{\prime}=0$ curves to demonstrate the fact that the stability regions in all cases are bounded by this heuristic necessary condition. In order to illustrate the sensitivity of the stability region to the value of the intrinsic frequency we have superimposed the results for $\omega=1$ and $\omega=10$ in Fig.~\ref{fig:fig3}(b,c,d). The darker colored solid regions show the stable regions for $\omega=1$ and the superposed lighter shaded regions pertain to $\omega=10$. We note that a higher intrinsic frequency $\omega$ expands the stable region. The results in Fig.~\ref{fig:fig3}(a) do not depend upon $\omega$. We also note that stable  traveling waves are possible even with zero delay beyond some critical value of $\kappa$ which in turn depends on the particular wave mode number. This is clearly seen in Fig.~\ref{fig:fig3}(b,c) for $k=1$ and $k=2$ respectively. 


The stability regimes of the $k=1$ and $k=2$ for $\omega=1$ have been superposed  in Fig.~\ref{fig:fig4}(a) highlighting the forbidden region in the inset. In Fig~(\ref{fig:fig4}(b,c,d)) we have further explored the dependency on $\omega$ by superposing 
the stability diagrams obtained for $\omega =0.8,0.8,0.5$ respectively. An interesting development is the constriction of the $k=1$ stability regime in the neck region seen in Fig~\ref{fig:fig4}(a)
to its breakup into two disconnected regions (in Fig.~\ref{fig:fig4}(b)) to the total disappearance of the stable region in the low $\kappa$ regime in Fig.~\ref{fig:fig4}(d). In Fig.~\ref{fig:fig4}(d) we also
notice disconnected stability regimes for $k=2$ and $k=3$. These regions exhibit a tongue structure for high connectivity (low $\kappa$) whereas they submerge into the stable region of the synchronous state for low connectivity (high $\kappa$).

\begin{figure}
\begin{tabular}
[c]{c}
\includegraphics[width=0.45\textwidth]{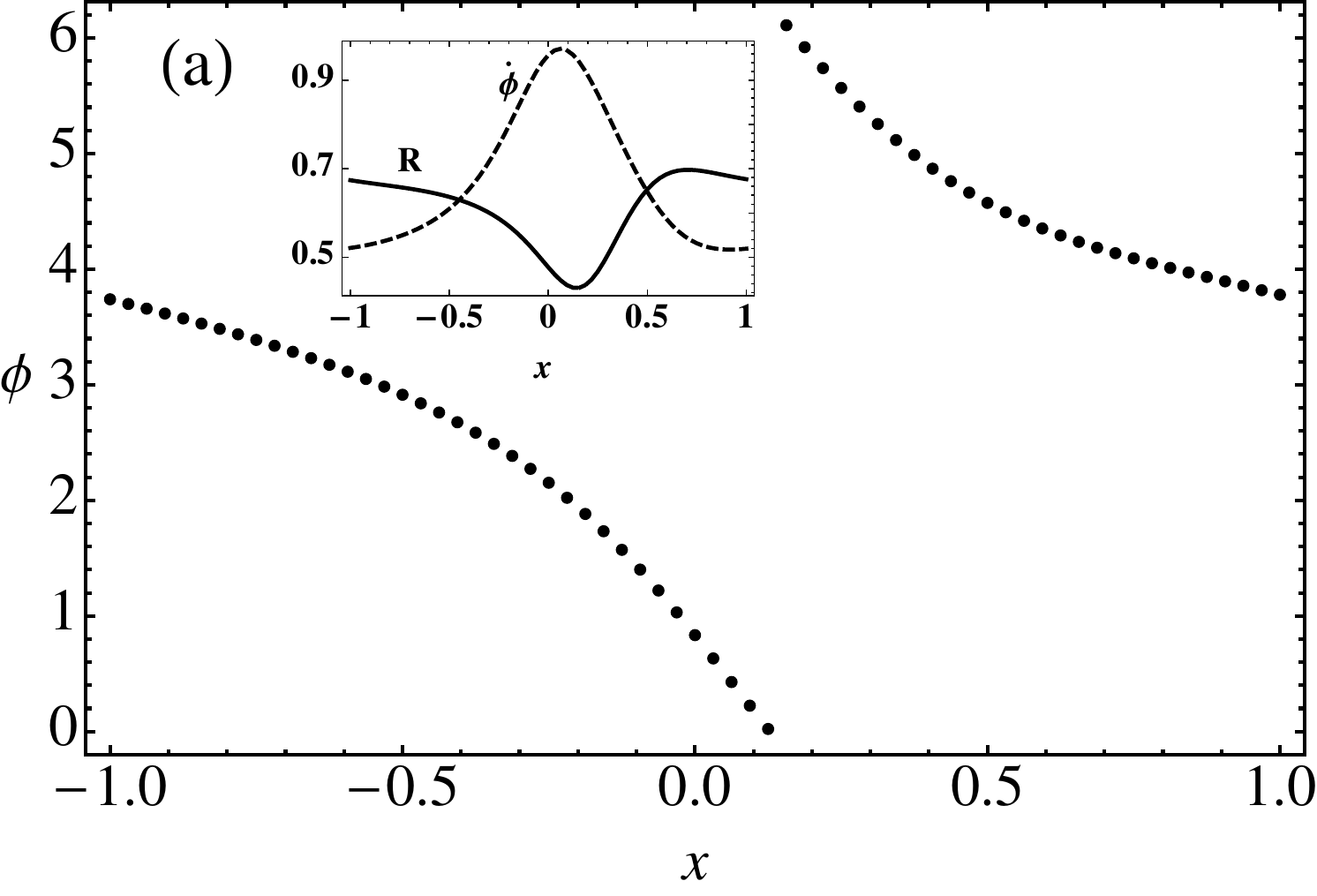}\\
\includegraphics[width=0.45\textwidth]{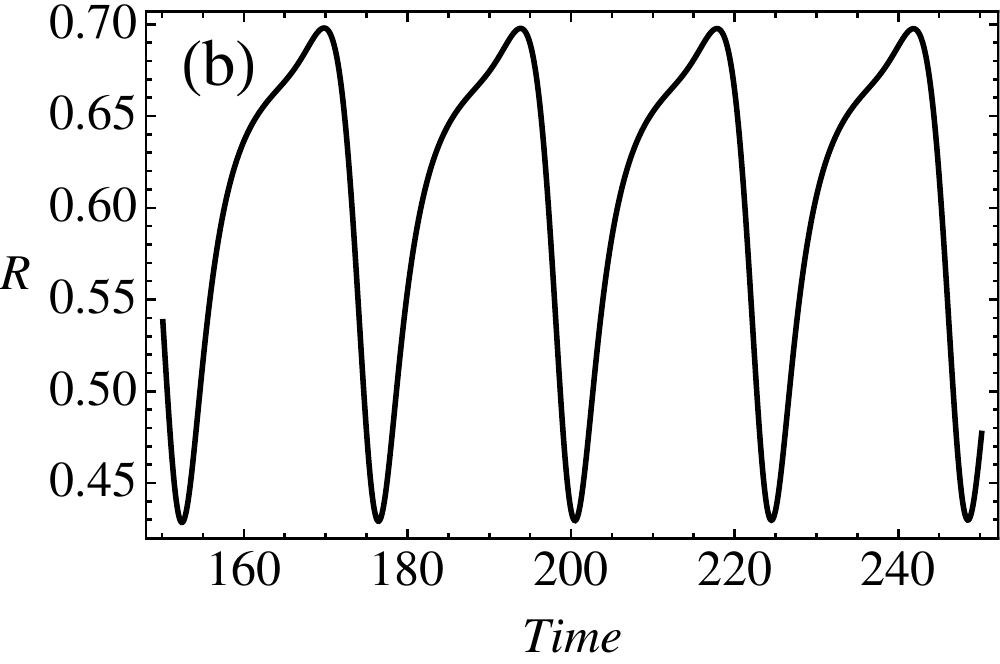}\\
\includegraphics[width=0.45\textwidth]{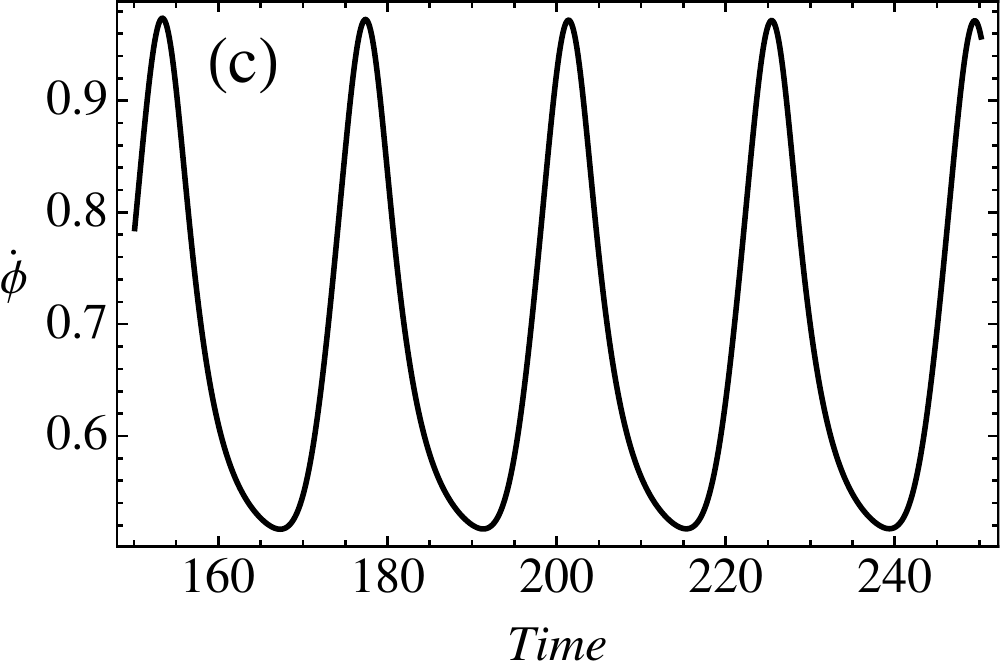}
\end{tabular}
\caption{  Spatio-temporal characteristics of a non-stationary breather state. Panel (a)  shows the snap shot of the spatial pattern of phase $\phi$ after the transients are over. The corresponding inset shows the spatial variations of phase velocity($\dot{\phi}$) and the amplitude $R$ of the complex order parameter ($\mathbb{Z}$).  The panels (b) \& (c) exhibit temporal behavior of $R$ and $\dot{\phi}$ respectively. The  parameter values are : $\omega=1$, $\kappa=2$ and $\tau_{m}=6.4$. }
\label{fig:fig5}
\end{figure}
\begin{figure}
\begin{tabular}
[c]{c}
\includegraphics[width=0.45\textwidth]{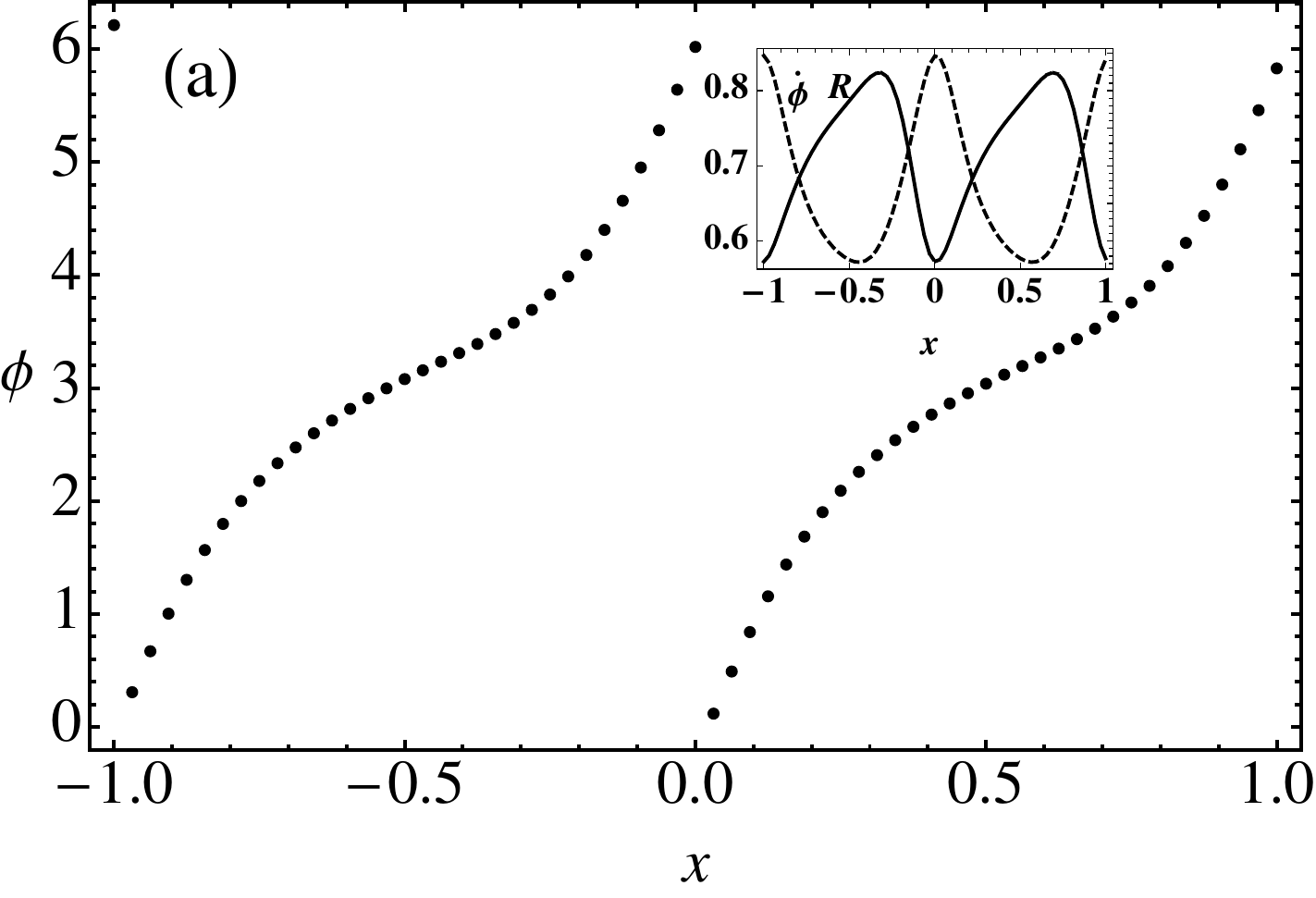}\\
\includegraphics[width=0.45\textwidth]{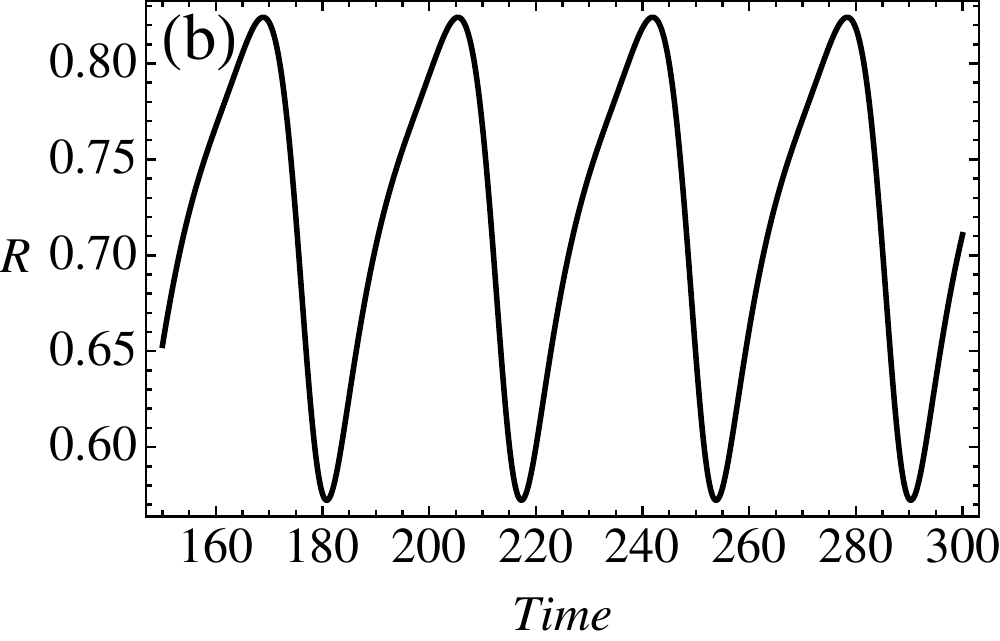}\\
\includegraphics[width=0.45\textwidth]{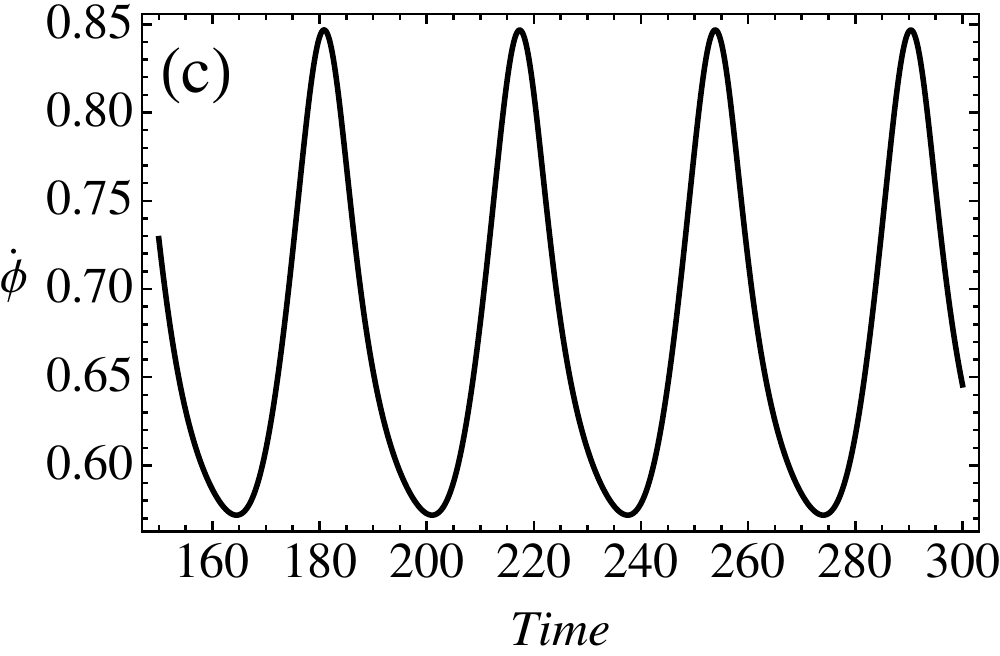}
\end{tabular}
\caption{The same as in Fig.~\ref{fig:fig5} but with parameter values:  $\omega=1$, $\kappa=10$ and $\tau_{m}=12.8$. }
\label{fig:fig6}
\end{figure}
An interesting feature of the stability diagram is the existence of forbidden regions where both the synchronous as well the traveling wave states are unstable indicating a forbidden zone in parametric space where no stable phase-locked modes can be sustained. This is first shown in Fig.~\ref{fig:fig2}(b) and subsequently in more details in Fig.~\ref{fig:fig4}(a-d). 
\section{Numerical Simulation of the model}
As an additional check of our stability results we have also carried out detailed and extensive numerical simulations of the discretized version of  Eq.~(\ref{phase}) and confirmed the existence 
of the stable and unstable regions shown in the various figures. In the course of these simulations we have also discovered the existence of a new class of  non-stationary solutions of the system that have not been observed before. For these states the phase $\phi(x,t)$ and its temporal derivative $\dot{\phi}(x,t)$ have a non-linear dependence on space and the order parameter of the system
has a spatial structure and also displays regular periodic variations in time.   Fig.~\ref{fig:fig5}(a)  shows a   snap shot of such a state with $\phi(x,t)$ as a function of $x$ and the corresponding inset shows the spatial 
structures of $\dot{\phi}(x,t)$ and the amplitudes $R$ of the complex order parameter $\mathbb{Z}$ defined as,
\begin{equation}
\mathbb{Z}(x,t)\equiv R e^{i \Phi}=\int_{-1}^{1}G(z)e^{i\left[\phi(x-z,t-|z|\tau_{m})\right]  }\,dz
\label{order_p}
\end{equation}
$\mathbb{Z}$ is a useful measure of the phase coherence of the oscillators in the system. Its amplitude $0\le R \le 1$ measures the system's coherence and $\Phi$ is the average phase.
%


Fig.~\ref{fig:fig5}(b,c) show the periodic temporal behavior of $R$ and $\dot{\phi}$ respectively  for this novel non-stationary {\it breather} state of the system. Fig.~\ref{fig:fig6}(a,b,c) show another non-stationary {\it breather} state for different set of system parameters. 
Note that the spatial characteristics of the non-stationary state are intermediate 
between a traveling wave state (which would have a linear dependence of the phase 
on distance) to a chimera state (whose phase vs space curve has a nonlinear nature 
that is broken up with regions of incoherent regions). The non-stationary solutions
have a smooth nonlinear curve for the spatial dependence of the phase (as shown in 
Fig.5(a) and Fig.6(a)). Its phase velocity (at any given spatial location) is a periodic function in time (as shown in Fig. 5(c) and 6(c)) in contrast to the 
traveling wave which has a constant phase velocity and the chimera state whose 
coherent portion also has a constant phase velocity.
We believe these non-stationary
{\it breather} states to be the intermediate link between the stationary traveling wave states and the clustered {\it chimera} states observed in \cite{sethia08}, in close
analogy to the evolutionary sequence observed in no-delay systems between synchronous states, breathers and chimeras \cite{abrams08}. Indeed as we move
away from the forbidden regions into the stable regimes for the traveling wave states we also find the co-existence of  {\it clustered} chimera states in agreement
with previous findings in \cite{sethia08}.\\

\section{Conclusions}
To conclude, we have carried out a detailed stability analysis of traveling wave solutions for a generalized Kuramoto model of coupled identical phase oscillators that
includes a spatially decaying coupling kernel and a space dependent time delay parameter. The comprehensive stability diagram presents a rich mosaic of multi-stable 
regions interspersed with unstable forbidden regions in the parametric space of the normalized quantities representing time delay $\bar{\tau}$, non-locality scale length $\kappa$
and the intrinsic frequency of an individual oscillator $\omega$. The present study complements our previous work on synchronous states to provide a consolidated stability diagram 
of phase locked states for the generalized Kuramoto model. It also ties up some of earlier equilibrium and stability results obtained from reduced models or simple limits of the present model to provide a more complete picture. Our investigations have also brought out important new findings such as the existence of forbidden  regions and the possibility of  exciting stable traveling waves  even with zero delay beyond some critical value of $\kappa$. Likewise the non-stationary ${\it breather}$ states are a novel find that provides further interesting information about
the collective dynamics of this system and in conjunction with our stability results on traveling waves and synchronous states may prove useful in the prediction and interpretation
of pattern formation phenomena in diverse practical applications.

\bibliographystyle{unsrt}



\end{document}